\documentclass[aps,prd,twocolumn,floatfix,superscriptaddress,nofootinbib,showpacs,
showkeys,preprintnumbers]{revtex4}
\usepackage{graphicx,amsmath}

\providecommand{\Journal}[4] {#1 {\bf #2}, #3 (#4)}
\providecommand{\EPJA}{Eur. Phys. J. A } %
\providecommand{\EPJC}{Eur. Phys. J. C } %
\providecommand{\CPC}{Comput. Phys. Commun. } %
\providecommand{\JPG}{J. Phys. G } %
\providecommand{\NIMA}{Nucl. Instr. Meth. A } %
\providecommand{\NPA}{Nucl. Phys. A } %
\providecommand{\NPB}{Nucl. Phys. B } %

\providecommand{\PAN}{Phys. At. Nucl. } %
\providecommand{\PLB}{Phys. Lett. B } %
\providecommand{\PRL}{Phys. Rev. Lett. } %
\providecommand{\PRC}{Phys. Rev. C } %
\providecommand{\PRD}{Phys. Rev. D } %
\providecommand{\YF}{Yad. Fiz. } %
\providecommand{\ZPA}{Z. Phys. A } %
\providecommand{\ibid}{{\it ibid. }} %

\newcommand{\pythia}{{\sc Pythia6}}

\begin{document}


\title{Search for an exotic $\mathbf{S=-2,\; Q=-2}$ baryon resonance at a mass
near 1862\,MeV in quasi-real photoproduction }

\def\groupalberta{\affiliation{Department of Physics, University of Alberta, Edmonton, Alberta T6G 2J1, Canada}}
\def\groupargonne{\affiliation{Physics Division, Argonne National Laboratory, Argonne, Illinois 60439-4843, USA}}
\def\groupbari{\affiliation{Istituto Nazionale di Fisica Nucleare, Sezione di Bari, 70124 Bari, Italy}}
\def\groupbeijing{\affiliation{School of Physics, Peking University, Beijing 100871, China}}
\def\groupchina{\affiliation{Department of Modern Physics, University of Science and Technology of China, Hefei, Anhui 230026, China}}
\def\groupcolorado{\affiliation{Nuclear Physics Laboratory, University of Colorado, Boulder, Colorado 80309-0446, USA}}
\def\groupdesy{\affiliation{DESY, 22603 Hamburg, Germany}}
\def\groupzeuthen{\affiliation{DESY, 15738 Zeuthen, Germany}}
\def\groupdubna{\affiliation{Joint Institute for Nuclear Research, 141980 Dubna, Russia}}
\def\grouperlangen{\affiliation{Physikalisches Institut, Universit\"at Erlangen-N\"urnberg, 91058 Erlangen, Germany}}
\def\groupferrara{\affiliation{Istituto Nazionale di Fisica Nucleare, Sezione di Ferrara and Dipartimento di Fisica, Universit\`a di Ferrara, 44100 Ferrara, Italy}}
\def\groupfrascati{\affiliation{Istituto Nazionale di Fisica Nucleare, Laboratori Nazionali di Frascati, 00044 Frascati, Italy}}
\def\groupgent{\affiliation{Department of Subatomic and Radiation Physics, University of Gent, 9000 Gent, Belgium}}
\def\groupgiessen{\affiliation{Physikalisches Institut, Universit\"at Gie{\ss}en, 35392 Gie{\ss}en, Germany}}
\def\groupglasgow{\affiliation{Department of Physics and Astronomy, University of Glasgow, Glasgow G12 8QQ, United Kingdom}}
\def\groupillinois{\affiliation{Department of Physics, University of Illinois, Urbana, Illinois 61801-3080, USA}}
\def\groupmit{\affiliation{Laboratory for Nuclear Science, Massachusetts Institute of Technology, Cambridge, Massachusetts 02139, USA}}
\def\groupmichigan{\affiliation{Randall Laboratory of Physics, University of Michigan, Ann Arbor, Michigan 48109-1120, USA }}
\def\groupmoscow{\affiliation{Lebedev Physical Institute, 117924 Moscow, Russia}}
\def\groupnikhef{\affiliation{Nationaal Instituut voor Kernfysica en Hoge-Energiefysica (NIKHEF), 1009 DB Amsterdam, The Netherlands}}
\def\groupstpetersburg{\affiliation{Petersburg Nuclear Physics Institute, St. Petersburg, Gatchina, 188350 Russia}}
\def\groupprotvino{\affiliation{Institute for High Energy Physics, Protvino, Moscow region, 142281 Russia}}
\def\groupregensburg{\affiliation{Institut f\"ur Theoretische Physik, Universit\"at Regensburg, 93040 Regensburg, Germany}}
\def\grouprome{\affiliation{Istituto Nazionale di Fisica Nucleare, Sezione Roma 1, Gruppo Sanit\`a and Physics Laboratory, Istituto Superiore di Sanit\`a, 00161 Roma, Italy}}
\def\groupsimonfraser{\affiliation{Department of Physics, Simon Fraser University, Burnaby, British Columbia V5A 1S6, Canada}}
\def\grouptriumf{\affiliation{TRIUMF, Vancouver, British Columbia V6T 2A3, Canada}}
\def\grouptokyo{\affiliation{Department of Physics, Tokyo Institute of Technology, Tokyo 152, Japan}}
\def\groupamsterdam{\affiliation{Department of Physics and Astronomy, Vrije Universiteit, 1081 HV Amsterdam, The Netherlands}}
\def\groupwarsaw{\affiliation{Andrzej Soltan Institute for Nuclear Studies, 00-689 Warsaw, Poland}}
\def\groupyerevan{\affiliation{Yerevan Physics Institute, 375036 Yerevan, Armenia}}


\groupalberta
\groupargonne
\groupbari
\groupbeijing
\groupchina
\groupcolorado
\groupdesy
\groupzeuthen
\groupdubna
\grouperlangen
\groupferrara
\groupfrascati
\groupgent
\groupgiessen
\groupglasgow
\groupillinois
\groupmit
\groupmichigan
\groupmoscow
\groupnikhef
\groupstpetersburg
\groupprotvino
\groupregensburg
\grouprome
\groupsimonfraser
\grouptriumf
\grouptokyo
\groupamsterdam
\groupwarsaw
\groupyerevan


\author{A.~Airapetian}  \groupmichigan
\author{N.~Akopov}  \groupyerevan
\author{Z.~Akopov}  \groupyerevan
\author{M.~Amarian}  \groupzeuthen \groupyerevan
\author{A.~Andrus}  \groupillinois
\author{E.C.~Aschenauer}  \groupzeuthen
\author{W.~Augustyniak}  \groupwarsaw
\author{R.~Avakian}  \groupyerevan
\author{A.~Avetissian}  \groupyerevan
\author{E.~Avetissian}  \groupfrascati
% ...

\collaboration{The HERMES Collaboration} \noaffiliation

\date{\today}

\begin{abstract}

A search for an exotic baryon resonance with $S=-2,\; Q=-2$ has been
performed in quasi-real photoproduction on a deuterium target
through the decay channel $\Xi^- \pi^- \to \Lambda\, \pi^- \pi^- \to
p \pi^- \pi^- \pi^-$. No evidence for a previously reported
$\Xi^{--}(1860)$ resonance is found in the $\Xi^- \pi^-$~invariant
mass spectrum. An upper limit for the photoproduction cross section
of 2.1\,nb is found at the 90\% confidence level. The
photoproduction cross section for the $\Xi^{0}(1530)$ is found to be
between $9$ and $24$\,nb.

\end{abstract}



\maketitle

The prediction for the existence of narrow exotic baryon
resonances~\cite{Dia97}, based on the Chiral Soliton Model, has
triggered an intensive search for the exotic members of an
anti-decuplet with spin 1/2. In this
anti-decuplet~\cite{Che85,Wal92,Dia97} all three vertices are
manifestly exotic. The lightest exotic member of this anti-decuplet
lying at its apex, named the $\Theta^+$,  was predicted~\cite{Dia97}
to have a mass of 1530\,MeV and a narrow width. It corresponds to a
$uudd\bar{s}$ configuration, and decays through the channels
$\Theta^{+}\rightarrow pK^0$ or $\Theta^{+}\rightarrow nK^{+}$.

Experimental evidence for the $\Theta^{+}$ first came from the
observation of a narrow resonance at $1540\pm 10\mbox{(syst)}$\,MeV
in the $K^-$ missing mass spectrum for the $\gamma n \to K^+K^-n$
reaction on $^{12}$C~\cite{SPring8}. The decay mode corresponds to a
$S$=$+1$ resonance and signals an exotic pentaquark state with quark
content ($uudd\bar{s}$). Further evidence emerged from a series of
experiments, with the observation of narrow
peaks~\cite{DIANA,JLab,SAPHIR,Asr03,JLabp,HERMES,ZEUS,SVD,COSY-TOF,JINR,LPI}
in the $nK^+$ and $pK^0_S$ invariant mass spectra near 1530\,MeV, in
most cases with a width that is limited by the experimental
resolution. Some doubts have been raised recently, however, because
of potential experimental artefacts~\cite{Dzierba} and the failure
to observe a signal in other
experiments~\cite{null-theta,BES,PHENIX,HERA-B,SPHINX,ALEPH,HyperCP}.

Experimental evidence for a second exotic member of the
anti-decuplet came from the reported observation of a $S$=$-2$,
$Q$=$-2$ baryon resonance in proton-proton collisions at
$\sqrt{s}=17.2$\,GeV at the CERN SPS~\cite{Alt03}. A narrow peak at
a mass of about $1862$\,MeV in the $\Xi^-\pi^-$ invariant mass
spectrum is proposed as a candidate for the predicted exotic
$\Xi^{--}_{3/2}$ baryon with $S$=$-2$, $I$=${\frac{3}{2}}$ and a
quark content of ($ddss\bar{u}$). At the same mass, a peak is
observed that is a candidate for the $\Xi^{0}_{3/2}$ member of this
isospin quartet. The corresponding anti-baryon spectra show
enhancements at the same invariant mass. The observed mass of
$1862$\,MeV falls below the prediction of Ref.~\cite{Dia97} and
above a prediction of Ref.~\cite{JW03}, although closer to the
latter. However, the result of Ref.~\cite{Alt03} has been
disputed~\cite{FW04}. In addition, this resonance has not been
confirmed by other experimental
searches~\cite{WA89,ALEPH,HERA-B-Xi}. Many further searches for the
$\Xi^{--}$ are presently underway~\cite{Xi-search} for which no
final results are available yet.

This paper presents the results of a search for the $\Xi^{--}$ in
quasi-real photoproduction on deuterium. The data were obtained by
the HERMES experiment with the 27.6\,GeV positron beam of the HERA
storage ring at DESY. Stored beam currents ranged from 9 to 45\,mA.
An integrated luminosity of 296\,pb$^{-1}$ was collected on a
deuterium gas target. The target was either longitudinally polarized
(223\,pb$^{-1}$) or unpolarized (73\,pb$^{-1}$). The yields were
summed over two spin orientations when the target was polarized.

The HERMES spectrometer~\cite{SPE} consists of two identical halves
located above and below the positron beam pipe, and has an angular
acceptance of $\pm 170$\,mrad horizontally, and $\pm$(40 --
140)\,mrad vertically. The trigger was formed by either a
coincidence between scintillator hodoscopes, a preshower detector
and a lead-glass calorimeter, or a coincidence between three
scintillator hodoscopes and two tracking planes, requiring that at
least one charged track appears in each of the detector halves of
the spectrometer.

The analysis searched for inclusive photoproduction of a
$\Xi^{--}$ followed by the decay $\Xi^{--} \to \Xi^- \pi^- \to
\Lambda\, \pi^- \pi^- \to p \pi^- \pi^- \pi^-$ or a $\Xi^{0} \to
\Xi^- \pi^+ \to \Lambda\, \pi^- \pi^+ \to p \pi^- \pi^- \pi^+$.
Events selected contained at least four tracks: three charged
pions in coincidence with one proton. Identification of charged
pions and protons was accomplished with a Ring-Imaging
\v{C}erenkov (RICH) detector~\cite{RICH} which provides separation
of pions, kaons and protons over most of the kinematic acceptance
of the spectrometer. The protons were restricted to a momentum
range of 2--15\,GeV/c and pions to a range of 0.25--15\,GeV/c.

Based on the intrinsic tracking resolution, the required event
topology included a minimum distance of approach between a proton
and a negative pion track less than 1.5\,cm (the midpoint of which
is defined as the $\Lambda$ decay vertex), a minimum distance of
approach between a second negative pion and the reconstructed
$\Lambda$ track less than 1.0\,cm (the midpoint of which is defined
as the $\Xi^- $ decay vertex), a minimum distance of approach
between a third pion and the reconstructed $\Xi^-$ track less than
2.5\,cm (the midpoint of which is defined as the production vertex),
a radial distance of the production vertex from the positron beam
axis less than 6\,mm, a $z$ coordinate of the production vertex in
the $\pm 20$\,cm long target cell within $-18$\,cm $< z < +18\,$cm
along the beam direction, a $\Lambda$ decay length (separation of
$\Xi^-$ and $\Lambda$ decay vertices) greater than 7\,cm, and a
$\Xi^-$ decay length (separation of production and $\Xi^-$ decay
vertices) greater than 10\,cm. Both possible $\pi^+\pi^-$
combinations were examined for a $K^0_S$ peak in the $M_{\pi^+
\pi^-}$ mass spectrum, and none were found.

The first step in the analysis was to search for $\Lambda$
candidates. The resulting invariant $M_{p\pi^-}$ spectrum is shown
in Fig.~1a. The position of the observed $\Lambda$ peak at
$1115.73\pm 0.01\mbox{(stat)}$\,MeV is in good agreement with the
nominal value~\cite{PDG}. Events were selected with a $M_{p\pi^-}$
invariant mass within $\pm 3\,\sigma$ of the centroid of the
$\Lambda$ peak. They were then combined with a $\pi^-$ to form the
$\Xi^-$ candidates. The resulting invariant $M_{\Lambda\pi^-}$
spectrum, shown in Fig.~1b, yields a $\Xi^-$ peak at $1321.8\pm
0.3\mbox{(stat)}$\,MeV, which is in good agreement with the expected
value of $1321.3$\,MeV~\cite{PDG}.

\begin{figure} [htb]
\begin{center}
\includegraphics[width=42mm,angle=0]{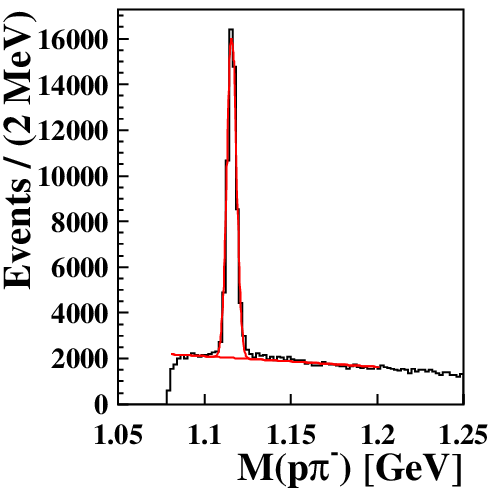}
\includegraphics[width=42mm,angle=0]{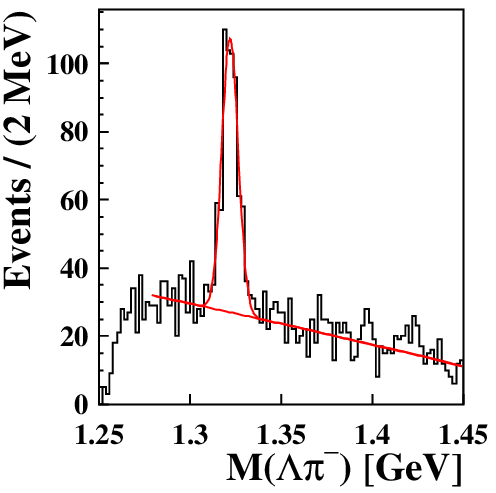}
\end{center}
\vspace{-5mm} \caption{(a) Invariant mass distribution of the
$p\pi^-$ system. (b) Invariant mass distribution of the
$p\pi^-\pi^-$ system. Both invariant mass distributions contain also
the charge conjugate (c.c.) modes ($\overline{p}\pi^+$,
$\overline{p}\pi^+\pi^+$), and are subject to the constraints in
event topology discussed in the text.\\ }
\label{fig1}
\end{figure}

The next step was to search for $\Xi^{--}_{3/2}$ ($\Xi^{0}_{3/2}$)
candidates in the $M_{\Xi^-\pi^-}$ ($M_{\Xi^-\pi^+}$) spectra.
Events were selected with a $M_{\Lambda\pi^-}$ invariant mass within
$\pm 3\,\sigma$ of the centroid of the $\Xi^-$ peak. The resulting
spectrum of the invariant mass of the $p\pi^-\pi^-\pi^-$ system is
displayed in Fig.~2. No peak structure is observed near 1862\,MeV.
Fig.~3 shows the resulting spectrum of the invariant mass of the
$p\pi^-\pi^-\pi^+$ system. While no peak structure is observed near
1862\,MeV, one appears at the mass of the known $\Xi^0(1530)$.

\begin{figure} [htb]
\begin{center}
\includegraphics[width=\linewidth,angle=0]{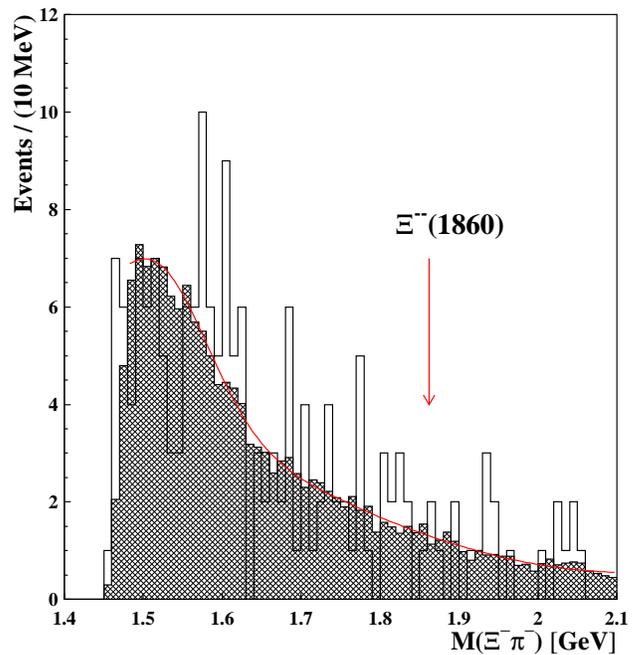}
\end{center}
\vspace{-5mm} \caption{Invariant mass distribution of the
$p\pi^-\pi^-\pi^-$ system, subject to the constraints in event
topology discussed in the text. The mixed-event background is
represented by the gray shaded histogram, which is normalized to
the background component of the fitted curve described in the
text. The arrow shows the hypothetical $\Xi^{--}_{3/2}$ mass.
} \label{fig2}
\end{figure}

The shape of the background  in Fig. 2 (Fig. 3) was determined from
randomly mixing $\Xi^-$ events with $\pi^-$ ($\pi^+$) tracks from
other events. Each resulting distribution was fit with a broad
Gaussian shape plus a polynomial. The $M_{\Xi^- \pi^-}$ distribution
shown in Fig.~2 was fit (smooth curve) with this background form
with a free normalization, together with a Gaussian shape whose
position was fixed at 1862\,MeV and whose width was fixed to the
instrumental resolution of $\sigma=10.2$\,MeV, derived from a
simulation of the spectrometer described below. The upper limit at
the 90\% confidence level (C.L.) for the hypothetical
$\Xi^{--}_{3/2}$ peak is found to be $3.9$ events. The $M_{\Xi^-
\pi^+}$  distribution was fit (curve in Fig.~3) using two peak
shapes plus the functional form that represents the mixed-event
background. One peak shape was Gaussian with a position fixed at
1862\,MeV and a fixed width of $\sigma=10.2$\,MeV, while the other
was a convolution of Gaussian and Breit-Wigner shapes that was
allowed to vary freely near the position of the well established
$\Xi^0(1530)$. This Gaussian contribution had a width fixed at
$7.2$\,MeV, while the Breit-Wigner shape had the width fixed at the
known $\Gamma = 9.1$\,MeV for the $\Xi^0(1530)$~\cite{PDG}. The
position of the $\Xi^0(1530)$ at $1536.5\pm 3.6\mbox{(stat)}$\,MeV
is consistent with the expected value of $1531.8$\,MeV. The values
for both Gaussian widths correspond to the simulated instrumental
resolutions. The area of the $\Xi^0(1530)$ peak is found to be
$35\pm 11$ events. For the hypothetical $\Xi^{0}_{3/2}$ peak at
1862\,MeV, the 90\% C.L. upper limit is found to be $4.6$ events.
All of these results are from unbinned maximum likelihood
fits~\cite{RooFit} to the original event distributions.

\begin{figure} [htb]
\begin{center}
\includegraphics[width=\linewidth,angle=0]{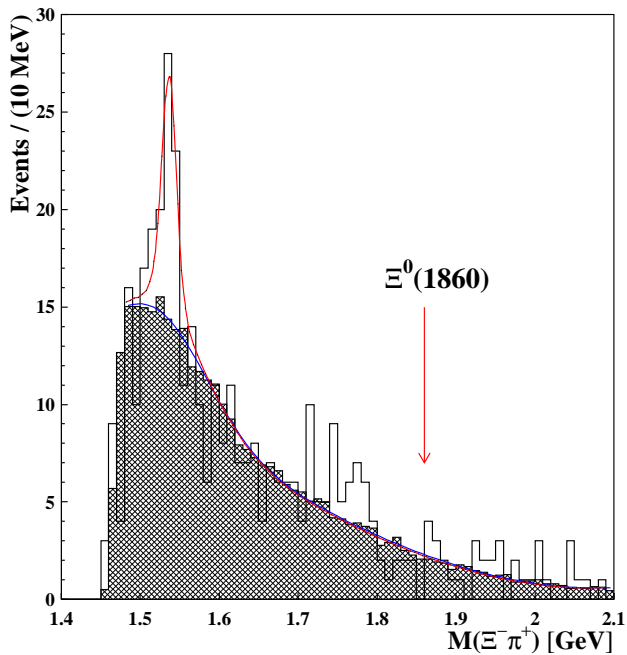}
\end{center}
\vspace{-5mm}
\caption{Invariant mass distribution of the
$p\pi^-\pi^-\pi^+$ (plus c.c.) system, subject to the constraints in
event topology discussed in the text. The mixed-event background is
represented by the gray shaded histogram, which is normalized to the
background component of the fitted curve described in the text. The
arrow shows the hypothetical $\Xi^{0}_{3/2}$ mass. The excess near
1.77\,GeV has a statistical significance of only $1.8\,\sigma$. }
\label{fig3}
\end{figure}

In order to derive cross sections, the detector acceptance was
simulated using two alternative models for the unknown production
kinematics. In the first ansatz, the unknown kinematic distribution
of the parent was described by functional forms resembling the
transverse and longitudinal momentum distributions of $\Lambda$
hyperons observed at HERMES. For the second model, the kinematic
distributions of the parent were taken to be those of the
$\Xi^0(1530)$ as predicted by the \pythia\ Monte Carlo
code~\cite{PYTHIA6}. Both models were isotropic in the decay angle
distribution.

In the case of the $\Xi^0(1530)$, the acceptance times efficiency
from these two approaches are 0.036\%\footnote{The acceptance for
the 4-track decay of the $\Xi^0(1530)$ resonance is comparable to
that for the 3-track decay of the $\Theta^+$~\cite{HERMES} due to
the inclusion of short tracks. These are tracks that make it
through the spectrometer magnet, but not all the way to the RICH
detector.} and 0.10\%, respectively. This range in acceptance
values due to the unknown production mechanism was carried forward
into the cross section calculations. The photoproduction cross
sections were evaluated using the Weizs\"acker-Williams flux
factor  calculated over the range $3.6\,\rm GeV < \nu < 24.6\,\rm
GeV$. If the branching ratio of $\Xi^0(1530) \to \Xi^- \pi^+$
decay is taken to be $2/3$~\cite{WA89-Xi1530}, its photoproduction
cross section is found to be between $8.8$ and $24$\,nb. This
experimental result is considerably higher than the \pythia\ model
prediction~\cite{PYTHIA6} in $4\pi$ acceptance of $0.92$\,nb for
production of $\Xi^0(1530)$ plus $\overline{\Xi}^0(1530)$.
However, the cross section ratio of $\overline{\Xi}^0(1530)$ to
$\Xi^0(1530)$ production predicted by \pythia\ is 1:4, which is
consistent with the present data.

To simulate the detector response for the hypothetical
$\Xi^{--}_{3/2}$ and $\Xi^{0}_{3/2}$ resonances, the particles were
generated with a mass of 1862\,MeV and an intrinsic width of
$\Gamma=2$\,MeV, at vertices distributed according to the HERMES
target gas profile. For the initial kinematic distributions, the
same two models as for the $\Xi^0(1530)$ were employed. The
instrumental resolution derived from this simulation was
$\sigma=10.2$\,MeV in the $M_{\Xi^-\pi^-}$ and $M_{\Xi^- \pi^+}$
mass spectra near 1860\,MeV.

Estimates of the spectrometer acceptance times efficiency from the
simulation were used to estimate an upper limit for the cross
section for production of the $\Xi^{--}$ and $\Xi^{0}$. Employing
the two models described above for the initial kinematic
distributions, the acceptances for these particles were estimated to
be 0.031\% and 0.065\%, respectively. Using the fit results
described above, the corresponding results for the upper limit for
the $\Xi^{--}$ ($\Xi^{0}$) photoproduction cross section times
branching fraction is found to be 1.0 to 2.1\,nb (1.2 to 2.5\,nb) at
the 90\% confidence level (C.L.). Based on the cross section ranges
for the $\Xi^0(1530)$ and $\Xi^{--}$ reported here, a 90\% C.L.
upper limit on the cross section ratio of $\Xi^{--} \to \Xi^-\pi^-$
to $\Xi^0(1530)$ between 0.06 and 0.15 has been derived. The values
were obtained by assuming a central value for each cross section
range, assigning the uncertainty to be half of the range, and
performing error propagation on the ratio.

In summary, a search for the exotic $\Xi^{--}$ baryon resonance has
been performed in quasi-real photoproduction on a deuterium target
through the decay channel $\Xi^- \pi^- \to \Lambda\, \pi^- \pi^- \to
p \pi^- \pi^- \pi^-$. The upper limit for the $\Xi^{--}$
photoproduction cross section is 2.1\,nb (90\% C.L.). In addition,
the well established $\Xi^0(1530)$ is clearly identified in this
experiment, and its photoproduction cross section is found to be
between $8.8$ and $ 24$\,nb, the range being due the effect of the
uncertainty in the production kinematics on the acceptance
correction.

\begin{acknowledgments}

We gratefully acknowledge the DESY management for its support and
the staff at DESY and the collaborating institutions for their
significant effort. This work was supported by the FWO-Flanders,
Belgium; the Natural Sciences and Engineering Research Council of
Canada; the National Natural Science Foundation of China; the
INTAS and ESOP network contributions from the European Community;
the German Bundesministerium f\"ur Bildung und Forschung (BMBF);
the Deutsche Forschungsgemeinschaft (DFG); the Deutscher
Akademischer Austauschdienst (DAAD); the Italian Istituto
Nazionale di Fisica Nucleare (INFN); Monbusho International
Scientific Research Program, JSPS, and Toray Science Foundation of
Japan; the Dutch Foundation for Fundamenteel Onderzoek der Materie
(FOM); the U. K. Engineering and Physical Sciences Research
Council and the Particle Physics and Astronomy Research Council;
and the U. S. Department of Energy and the National Science
Foundation.

\end{acknowledgments}

\end{document}